\documentclass[letterpaper,pra,twocolumn,notitlepage,superscriptaddress, nofootinbib, showpacs, eqsecnum]{revtex4-1}

\usepackage {amsmath}
\usepackage{graphicx}
\usepackage{braket}
\usepackage{amsthm}
\usepackage{amssymb}
\usepackage{color}
\usepackage{mathtools}
\usepackage{tikz}
\usetikzlibrary{decorations}
\usetikzlibrary{patterns}

\usepackage[colorlinks=true,urlcolor=blue, linkcolor=blue]{hyperref}
\usepackage{enumitem}

\providecommand{\abs}[1]{\lvert#1\rvert}

\def\({\left(}
\def\){\right)}

\def\eq#1{Eq.~(\ref{eq:#1})}

\def\fig#1{Fig.~\ref{fig:#1}}

\def\sec#1{Sec.~\ref{sec:#1}}

\newcommand{\blk}{\color{black}}

\DeclareMathOperator{\sech}{sech}

\DeclareMathOperator{\Imtext}{Im}

\renewcommand{\Im}{\Imtext}

\usepackage{bm}

\newcommand{\op}[1]{\hat{#1}}
\newcommand{\opvec}[1]{\op{\vec{#1}}}

\newcommand{\mat}[1]{\bm{\mathrm{#1}}}

\renewcommand{\vec}[1]{\bm{\mathrm{#1}}}

\newcommand{\tp}{\mathrm{T}}

\graphicspath{ {./Figures/} }



\setlist[description]{leftmargin=0pt}

\begin{document}
\title{Flexible quantum circuits using scalable continuous-variable cluster states}
%
%
%
%
\author{Rafael N. Alexander} \email{rafael.alexander@sydney.edu.au}
\affiliation{School of Physics, The University of Sydney, Sydney, NSW 2006, Australia}
\affiliation{School of Science, RMIT University, Melbourne, VIC 3001, Australia}
\author{Nicolas C. Menicucci} \email{ncmenicucci@gmail.com}
\affiliation{School of Physics, The University of Sydney, Sydney, NSW 2006, Australia}
\affiliation{School of Science, RMIT University, Melbourne, VIC 3001, Australia}
\date{\today}
\begin{abstract}
We show that measurement-based quantum computation on scalable continuous-variable~(CV) cluster states admits more quantum-circuit flexibility and 
compactness
than similar protocols for standard square-lattice CV cluster states. This advantage is a direct result of the macronode structure of these states---that is, a lattice structure in which each graph node actually consists of several physical modes. These extra modes provide additional measurement degrees of freedom at each graph location, which can be used to manipulate the flow and processing of quantum information more robustly and with additional flexibility that is not available on an ordinary lattice.%
\end{abstract}
\maketitle
%
\section{Introduction}\label{sec:introduction}
Quantum information processing using measurement-based quantum computing~(MBQC)%
~\cite{Raussendorf2001} is divided into two steps: (1)~preparation of a universal, highly-entangled resource state (the standard choice is a \emph{cluster state} with a square-lattice graph~\cite{Briegel2001}),  followed by (2)~a sequence of single-site projective measurements with feedforward. %

The last 15 years have seen the emergence of numerous extensions, improvements, and generalizations of this basic model.
Important for this work is its generalization from cluster states made of qubits to those made of continuous-variable~(CV) quantum systems~\cite{Menicucci2006}. 
Unlike their photonic-qubit counterparts~\cite{Nielsen2004, Browne2005}, optical CV cluster states can be generated both deterministically and on a large scale with minimal experimental equipment. They need only offline squeezing and linear optics~\cite{VanLoock2007a}, all of which can be implemented using a single optical parametric oscillator (OPO)~\cite{Menicucci2007, Menicucci2008, Flammia2009, Alexander2015}.
 Extremely large cluster states of this type can be made with existing technology based on either frequency modes~\cite{Chen2014,Wang2013} or temporal modes~\cite{Yokoyama2013,Menicucci2011a}. 

Using CV cluster states for quantum computation comes with a price. Ideal states are infinitely squeezed~\cite{Menicucci2006, Gu2009}; thus, noise is introduced into the computation due to the fact that only finite squeezing resources (and hence, finite energy) can be used in generating the state~\cite{Gu2009,Cable2010,Alexander2014}. If left unchecked, this noise limits the length of computation possible using these states~\cite{Ohliger2010,Ohliger2012}. Nevertheless, it is still possible to achieve universal fault-tolerant quantum computation with CV cluster states~\cite{Menicucci2014} by employing known quantum-error-correction protocols~\cite{Gottesman2001}, provided that the experimentally achievable squeezing levels are high enough. The current best recorded squeezing level in an optical setup is 12.7 dB of squeezing~\cite{Eberle2010}, whilst the lowest theoretical upper bound on the required squeezing for fault tolerant quantum computing is 20.5 dB~\cite{Menicucci2014}. 

Closing this squeezing gap in scalable CV cluster state implementations is of paramount importance for their use in large-scale,  fault-tolerant quantum computation. A significant step in this direction is the development of resource-customized measurement-based protocols that capitalize on the available squeezing in order to minimize the noise per gate~\cite{Alexander2014, Alexander2015, Ferrini2015a}. 

In the same vein, here we give a new measurement protocol that is customized for a type of universal CV cluster state that is particularly scalable, known as the \emph{quad-rail lattice} (QRL)~\cite{Menicucci2011a, Wang2013}. The generation procedure of the QRL is particularly simple owing to the fact that its graph~\cite{Menicucci2011a, Wang2013} is self-inverse and bipartite~\cite{Menicucci2008, Flammia2009}. Indeed, it needs only two-mode squeezed states (TMSSs) and a single 4-port linear optics gate (known as a foursplitter) as building blocks~\cite{Menicucci2011a, Wang2013}. This state's graph contains within it a square-lattice topology (making it universal) with respect to four-mode lattice sites known as \emph{macronodes}. Our protocol leverages extra degrees of freedom present in each macronode, resulting in improved circuit compactness and flexibility. This work extends the macronode protocol presented in Ref.~\cite{Alexander2014}, which applies to the 1D resource state known as the  \emph{CV dual-rail wire}~\cite{Menicucci2011a, Yokoyama2013, Wang2013, Chen2014, Alexander2014}. 

The structure of this Article is as follows: In \sec{background} we review some basics of Gaussian pure states and the QRL~\cite{Menicucci2011a, Wang2013}. In \sec{macrointro} we introduce the basic components of our measurement protocol, including encoding, unitary gates, and measurement readout. In \sec{flexcirc} we describe how these elements can be composed, allowing for flexible design of quantum circuits.  In \sec{comparison} we compare this protocol to previous work. We conclude in \sec{conc}.

%

\section{Background}\label{sec:background}

Throughout this Article, we adopt the following conventions for all modes: $\op q = \tfrac {1} {\sqrt 2} (\op a + \op a^\dag)$, $\op p = \tfrac {1} {i\sqrt 2} (\op a - \op a^\dag)$. Using $[\op a, \op a^\dag] = 1$, this implies that $[\op q, \op p] = i$ with $\hbar = 1$.

\subsection{Symplectic formalism and gate definitions}

The Heisenberg action of an $N$-mode Gaussian unitary~$\hat{U}$ acting on the vector of Heisenberg-picture operators ${\hat{\mathbf{x}}=\left( \begin{smallmatrix}\hat{\mathbf{q}} \\ \hat{\mathbf{p}} \end{smallmatrix} \right)}$ can be written as
\begin{align}
\label{eq:heisact}
	\hat{U}^\dag \hat{\mathbf{x}} \hat{U} = \mathbf{S}_{\op U} \hat{\mathbf{x}},
\end{align}
where we have ignored displacements and
\begin{align}
\label{eq:Usymp}
\mathbf{S}_{\op U} =\begin{pmatrix} \mathbf{A} & \mathbf{B} \\ \mathbf{C} & \mathbf{D} \end{pmatrix}
\end{align}
is a $2N\times 2N$ real, symplectic matrix. Some useful examples are given below.

\vspace{1em}\noindent \textbf{The phase-delay gate} is defined to be
\begin{align}
\label{eq:rotationgate}
 \hat{R}(\theta) &\coloneqq \exp(i \theta \op a^\dag \op a) \nonumber \\
	&= \exp{\left[ \frac{i \theta}{2} (\hat{q}^{2}+ \hat{p}^{2}-1)\right]}.
\end{align}
Its Heisenberg action on $\op{\mathbf{x}} = (\op{q}, \op{p})^\tp$ is given by the symplectic matrix
\begin{equation}
\mathbf{R}(\theta ) =\begin{pmatrix} \cos\theta & -\sin\theta \\ \sin\theta & \cos\theta \end{pmatrix}.
\end{equation}
Note that $\hat{R}(-\omega\, \delta t)$ implements \emph{forward} time evolution for an oscillator with frequency $\omega$ over a small time interval $\delta t>0$. Thus, for \emph{positive} $\theta$, the gate $\hat{R}(\theta)$ will \emph{delay} the oscillator by a time interval~$\theta/\omega$. This motivates our choice of terminology and sign convention for this gate. %

In the Schr\"odinger picture, a phase delay by~$\theta$ [i.e., $\hat{R}(\theta)$] rotates the state's Wigner function \emph{counter-clockwise} by an angle~$\theta$. Viewed instead from the Heisenberg picture, this operation rotates the vector~$\hat{\mathbf{x}}$ of quadrature operators in the same fashion---i.e., counter-clockwise by~$\theta$.

\vspace{1em}\noindent \textbf{The single-mode squeezing gate} we use has the following (nonstandard) definition:
\begin{align}
\hat{S}(s) &\coloneqq \hat{R}(\text{Im} \ln{s})\exp\left[-\frac 1 2 (\text{Re} \ln s) (\op a^2-\op a^{\dag2}) \right] \nonumber \\
	&= \hat{R}(\text{Im} \ln{s})\exp\left[-\frac i 2 (\text{Re} \ln s) (\hat{q}\hat{p}+\hat{p}\hat{q}) \right],\label{eq:squeezegate}
\end{align}
where $s\in\mathbb{R}\backslash \{0 \}$ is called the \emph{squeezing factor}. This is related to the more commonly used \emph{squeezing parameter}~$r$ through 
\begin{align}
\abs{s} = e^r.
\end{align}
This gate differs from the ordinary squeezing gate only by an additional $\pi$ phase delay when $s<0$. Its Heisenberg action on $\op{\mathbf{x}}$ is given by the symplectic matrix
\begin{equation}
\mathbf{S}(s) = \begin{pmatrix} s & 0 \\ 0 & s^{-1} \end{pmatrix}.
\end{equation}
In the Heisenberg picture, this evolution multiplies the $\op q$ quadrature by~$s$ and the $\hat{p}$ quadrature by~$s^{-1}$. We define this to be what is meant by ``squeezing by a factor of~$s$''. (In addition to the $\pi$ phase delay when $s<0$, this operation anti-squeezes~$\op q$ and squeezes~$\op p$ when $\abs{s}>1$, and vice versa if $\abs{s}<1$.) 

\vspace{1em}\noindent \textbf{The beamsplitter gate} is defined to be 
\begin{align}
	\hat{B}_{ij}(\theta)&\coloneqq \exp \left[-\theta (\op a_i^\dag \op a_j - \op a_j^\dag \op a_i) \right] \nonumber \\
	& = \exp[-i\theta(\hat{q}_{i}\hat{p}_{j}-\hat{q}_{j}\hat{p}_{i})],
\end{align}
where $\sin \theta$ is the reflectivity of the beamsplitter. %
Its Heisenberg action on $\op{\mathbf{x}}=(\op{q}_{i}, \op{q}_{j}, \op{p}_{i}, \op{p}_{j})^\tp$ is given by 
\begin{align}
	\mathbf{B}_{ij}(\theta) = \begin{pmatrix} \cos{\theta} & -\sin{\theta} & 0 & 0 \\ \sin{\theta} & \cos{\theta} &  0 & 0 \\ 0 & 0 & \cos{\theta} & -\sin{\theta} \\ 0 & 0 & \sin{\theta} & \cos{\theta} \end{pmatrix}.
\end{align}
While this beamsplitter is often defined in the literature with additional phase delays incorporated (in order to match the physics more closely), the definition here matches that in Refs.~\cite{Menicucci2011,Menicucci2011a,Alexander2014} and is more suitable for analysis of CV quantum-computing applications.

A useful property of this gate is that, up to displacements,  it commutes with the action of the same single-mode Gaussian unitary gate on two-modes, i.e., 
\begin{align}
\left[ \hat{B}_{ij} (\theta), \hat{U}_{i}\hat{U}_{j} \right] = 0, \label{eq:BScom}
\end{align}
where $\hat{U}$ is a single-mode Gaussian unitary gate without displacements. 
\begin{proof}[Proof of Equation~\eqref{eq:BScom}]
It suffices to check that their symplectic matrix representations commute. Denote the symplectic matix representation of $\hat{U}$ by $\mathbf{U}$. Then, the symplectic matrix representation of $\hat{U}_{i}\hat{U}_{j}$ can be represented as $\mathbf{U}\otimes\mathbf{I}$, where $\otimes$ is a kronecker product and $\mathbf{I}$ is the $2\times 2$ identity matrix.  Note similarly that $\mathbf{B} (\theta) = \mathbf{I}\otimes \mathbf{R}(\theta)$. Clearly, these matrices commute.  
\end{proof}

\vspace{1em}\noindent \textbf{The 50:50 beamsplitter gate} is defined as 
\begin{align}
\hat{B}_{ij} \coloneqq \hat{B}_{ij} \left(\frac{\pi}{4}\right), 
\end{align}
i.e., it is a special case of the above defined beamsplitter where $\theta=\frac{\pi}{4}$, and the dependence on the angle is dropped for notational convenience.  
Note that $\op{B}^{\dagger}_{ij} = \op{B}_{ji}$.

\begin{figure}
\includegraphics[width=1\columnwidth]{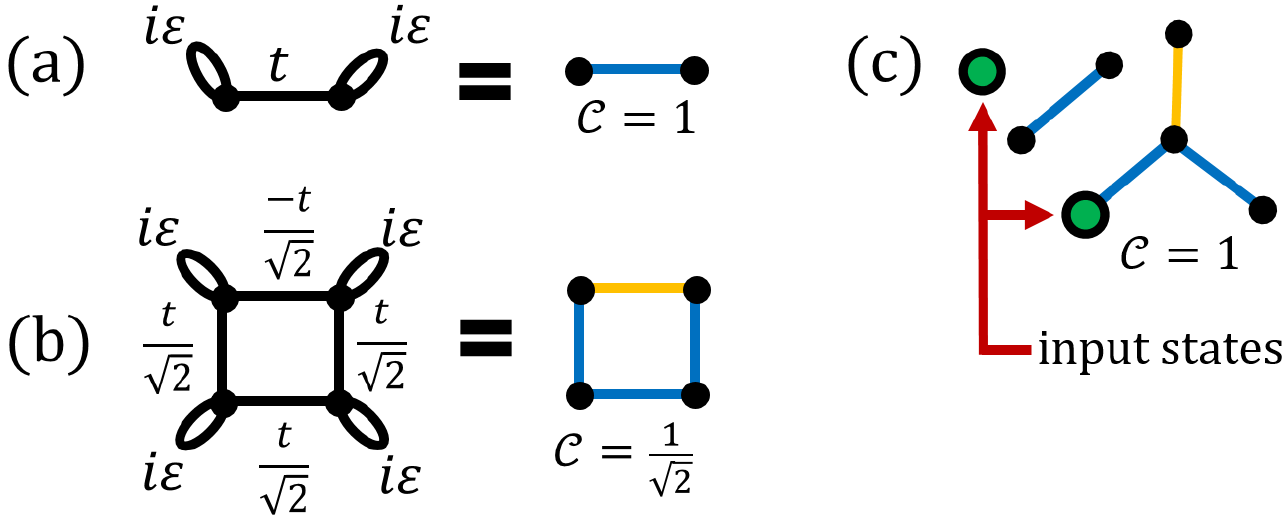}
\caption{\textbf{(a)}~Two-mode continuous-variable cluster state represented using the full graphical calculus~\cite{Menicucci2011} (left) and the simplified graphical calculus~\cite{Menicucci2011a} (right). Edge weights $\varepsilon$ and $t$ are defined in Eq.~\eqref{eq:epsilon} and Eq.~\eqref{eq:t}, respectively. \textbf{(b)}~We similarly represent a four-mode square CV cluster state. \textbf{(c)}~Seven-mode state containing two inputs (green nodes)---one is disconnected (tensor product with the rest of the state), and the other is attached to a three-mode Gaussian pure state. 
}\label{fig:simpgraphcalc}
\end{figure}

\vspace{1em}\noindent Finally, \textbf{the foursplitter gate} is defined to be 
 \begin{align}
&\op{A}_{j k l m}\coloneqq \exp\left[ \frac{\pi}{4}\left((\op{a}^{\dagger}_{k}+ \op{a}^{\dagger}_{l})(\op{a}_{j} - \op{a}_{m})- \text{h.c.}
\right)\right] \nonumber \\ & = \exp\left[- i \frac{\pi}{4}\Bigl((\op{q}_{j} - \op{q}_{m})(\op{p}_{k}+ \op{p}_{l})+(\op{q}_{k}+\op{q}_{l})(\op{p}_{m}-\op{p}_{j})\Bigr)\right], \label{eq:FS}
 \end{align}
where ``h.c." abbreviates the hermitian conjugate ($\dagger$) of the first term in the exponent. Its Heisenberg action on $\op{\mathbf{x}}=(\op{q}_{i}, \op{q}_{j}, \op{q}_{k}, \op{q}_{l}, \op{p}_{i}, \op{p}_{j}, \op{p}_{k}, \op{p}_{l})^\tp$ is given by 
\begin{align}
\mathbf{A}_{i j k l} \coloneqq \begin{pmatrix} \tilde{\mathbf{A}} & \mathbf{0} \\ \mathbf{0} & \tilde{\mathbf{A}}  \end{pmatrix},\label{eq:foursplit}
\end{align}
where $\mathbf{0}$ denotes the $4\times4$ matrix of zeroes and
\begin{align}
\tilde{\mathbf{A}}=\frac{1}{2}\begin{pmatrix} 1 & -1 & -1 & 1 \\ 1 & 1 & -1 & -1 \\ 1 & -1 & 1 & -1 \\ 1 & 1 & 1 & 1 \end{pmatrix}. \label{eq:FSsubmat}
\end{align}
This gate admits the following convenient decompositions into four 50:50 beamsplitters~\cite{Menicucci2011a,Wang2013}:
\begin{align}
	\op{A}_{ijkl}= \op{B}_{ij}  \op{B}_{kl} \op{B}_{ik}  \op{B}_{jl}= \op{B}_{ik}  \op{B}_{jl} \op{B}_{ij}  \op{B}_{kl}. \label{eq:bsdecomp}
\end{align}

\subsection{Graphical calculus for Gaussian pure states}
%

In this Article, we will be describing the properties of a Gaussian pure state (the QRL). For convenience, we will represent this state by its graph~\cite{Menicucci2011}, which is defined using the graphical calculus for Gaussian pure states, summarized below.

\emph{Graphs}.---%
%
Given an undirected, complex-weighted graph on $N$ nodes with adjacency matrix~$\mathbf{Z}$ (${= \mathbf{Z}^\tp}$)  and $\Im \mathbf Z > 0$%
~\cite{Menicucci2011}, $\mathbf{Z}$ uniquely defines the position-space wavefunction %
\begin{align}
	\psi_{\mathbf{Z}}(\mathbf{q}) \coloneqq \frac{( \det \Im \mat Z)^{1/4}}{\pi^{N/4}}\exp\left[\frac{i}{2}\mathbf{q}^\tp\mathbf{Z}\mathbf{q}\right]
\end{align}
of the $N$-mode Gaussian pure state $\ket{\psi_{\mathbf{Z}}}$, where $\mathbf{q}$ is a column vector of $c$-numbers. It also gives a compact description of the nullifiers of $\ket{\psi_{\mathbf{Z}}}$:
\begin{align}
	\left(\hat{\mathbf{p}} - \mathbf{Z}\hat{\mathbf{q}}\right)\ket{\psi_\mathbf{Z}}= \mathbf{0},
\end{align}
where $\hat{\mathbf{q}}=(\hat{q}_{1}, \dots \hat{q}_{N})^\tp$ and $\hat{\mathbf{p}}=(\hat{p}_{1}, \dots \hat{p}_{N})^\tp$ are column vectors of operators. Every Gaussian pure state uniquely defines (up to phase-space displacements and overall phase) an associated graph~$\mathbf{Z}$~\cite{Menicucci2011}. 

\emph{Graph update rule}.---%
In the language of the graphical calculus, Schr\"odinger-picture evolution of a Gaussian unitary $\hat{U}$ can be represented up to displacements and overall phase by a graph update rule
\begin{align}
\hat{U} \ket{\psi_{\mathbf{Z}}} = \ket{\psi_{\mathbf{Z}'}}
\end{align}
with
\begin{align}
\label{eq:graphtrans}
	\mathbf{Z}'=(\mathbf{C}+\mathbf{D}\mathbf{Z})(\mathbf{A}+\mathbf{B}\mathbf{Z})^{-1}
\end{align}
where the submatrices $\mathbf{A}$, $\mathbf{B}$, $\mathbf{C}$ and $\mathbf{D}$ are defined via the Heisenberg action of $\hat{U}$, as in Eq.~\eqref{eq:Usymp}.

\emph{Simplified graphs}.---%
The Gaussian pure states that we consider in this Article are specified by few graphical parameters, i.e., edge and self-loop weights in $\mathbf{Z}$. When representing such states by their corresponding graph, it is convenient to use a simplified set of rules known as the \emph{simplified graphical calculus}~\cite{Menicucci2011a}. It makes use of the following conventions: no self-loops are drawn, and the color of an edge indicates the sign of its edge weight. See \fig{simpgraphcalc}(a) and (b).  In addition to these (standard) conventions, we will use differently colored nodes---green instead of black---to denote the inclusion of an input state localised to a single graph node, as shown in \fig{simpgraphcalc}(c). The self-loops (not shown) on all non-input graph nodes have weight 
\begin{align}
 i  \varepsilon \coloneqq i\sech{2r},
\label{eq:epsilon}
\end{align}
where the squeezing parameter $r$ gives the amount of vacuum squeezing used in preparing the state~\cite{Menicucci2011a, Wang2013}. All edge weights between different nodes are 
\begin{align}
\pm \mathcal{C} t \coloneqq \pm \mathcal{C} \tanh{2r} , \label{eq:t}
\end{align}
which is the product of the \emph{edge-weight coefficient} $\mathcal{C}$ (specified on each figure) and a squeezing-dependent factor~$t$, along with a sign $\pm$ denoted by blue/yellow, respectively. Note that $\varepsilon \to 0$ and $t \to 1$ as the squeezing parameter~$r \to \infty$, which corresponds to the high-squeezing limit.

Although our use of the graphical calculus strictly applies only when the input states (green nodes) are themselves Gaussian pure states, this choice is purely for representational convenience. All results presented here hold for general input states, including non-Gaussian and/or mixed states. 
%

\subsection{The quad-rail lattice}
\begin{figure}[t!]
\includegraphics[width=\columnwidth]{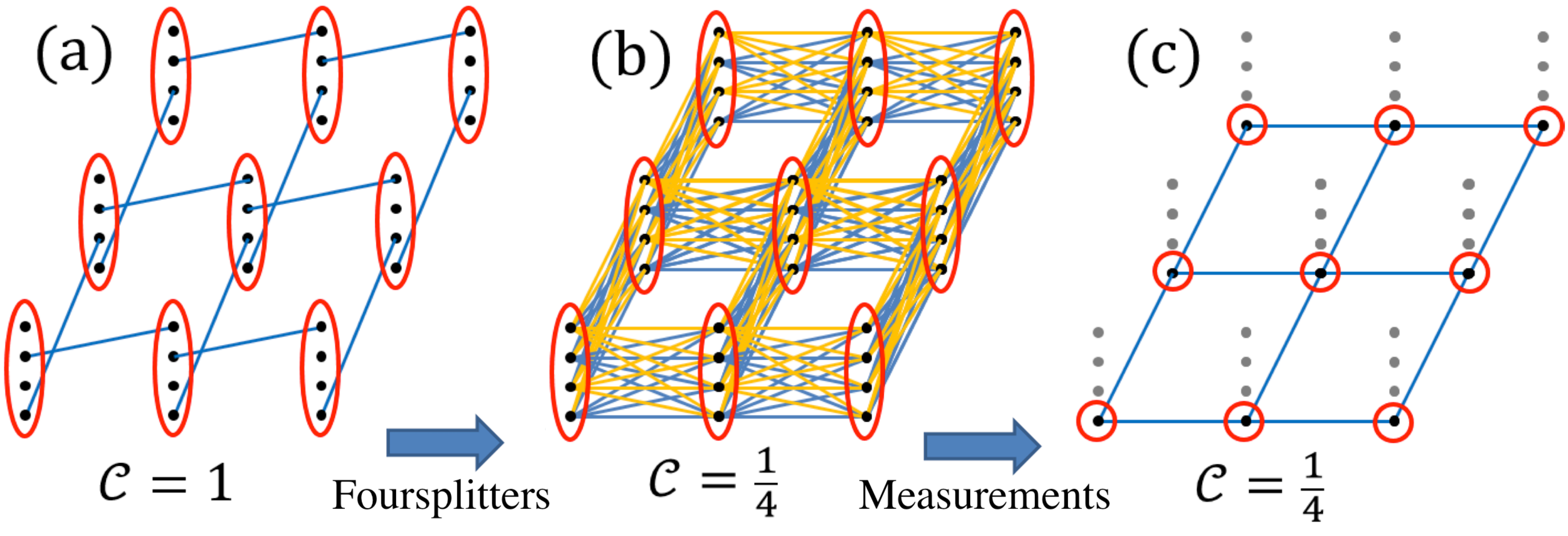}
\caption{Simplified graphical-calculus representation~\cite{Menicucci2011a} of the construction of the quad-rail lattice and conversion to an ordinary continuous-variable cluster state. \textbf{(a)}~A collection of two-mode continuous-variable cluster states. These pairs are ``stitched" together by a foursplitter gate (Eq.~\eqref{eq:FS}) at each macronode (indicated by the red ovals) in order to construct the quad-rail lattice. \textbf{(b)}  This graph defines the quad-rail lattice state. \textbf{(c)} Measuring the top three layers (faded) of the quad-rail lattice in the $\hat{q}$ basis produces a square-lattice continuous-variable cluster state as shown. Note that each site (red circle) only contains one mode. In the original proposal~\cite{Menicucci2011a}, universal quantum computation proceeded via the standard measurement-based protocol~\cite{Gu2009}. Removing the extra nodes and links from~(b), however, wastes squeezing resources~\cite{Alexander2014}. Instead, our proposal directly employs the state shown in~(b), making more efficient use of the available resources (the advantages are discussed in \sec{comparison}).%
}\label{fig:QRL}
\end{figure}

The QRL can be generated from a collection of two-mode cluster states [defined in Fig.~\ref{fig:simpgraphcalc}(a)]\footnote{Equivalently, two-mode squeezed states can be used by incorporating a $\frac{\pi}{4}$ phase delay into the measurement of all nodes~\cite{Gabay2016}.} arranged along edges of a square lattice by applying a foursplitter gate [Eq.~\eqref{eq:FS}] to each four-mode lattice site, a.k.a. a  \emph{macronode}~\cite{Menicucci2011a,Wang2013}.  The resulting QRL is defined by its four-layered square-lattice graph, as shown in \fig{QRL}(b). Further details about the generation of this state can be found in Refs.~\cite{Menicucci2011a,Wang2013}.

The QRL is universal for MBQC. To see this, consider measuring the top three layers of modes in $\hat{q}$ basis. Note that such measurements can be implemented experimentally via homodyne detection~\cite{Menicucci2011a, Wang2013}. Graphically, this action is represented by node deletion~\cite{Menicucci2011}, resulting in square-lattice CV cluster state as shown in \fig{QRL}(c).  Up to displacements, this is the canonical resource state for universal MBQC with CVs~\cite{Menicucci2006, Gu2009}. Unwanted displacements (due to $\hat{q}$ measurements on the top three layers) can be straightforwardly taken into account in the measurement protocol by feedforward. 

Achieving universal quantum computation this way is not optimal, however, because projecting down to a canonical CV cluster state %
results in an ordinary lattice with ${\mathcal C = \tfrac 1 4}$ (instead of ${\mathcal C=1}$), which introduces excessive noise when used in a computation~\cite{Alexander2014}.
We use the remainder of this Article to introduce a different---and much more favorable---MBQC protocol that runs directly on the full QRL, \fig{QRL}(b).

\begin{figure}
\includegraphics[width=1\columnwidth]{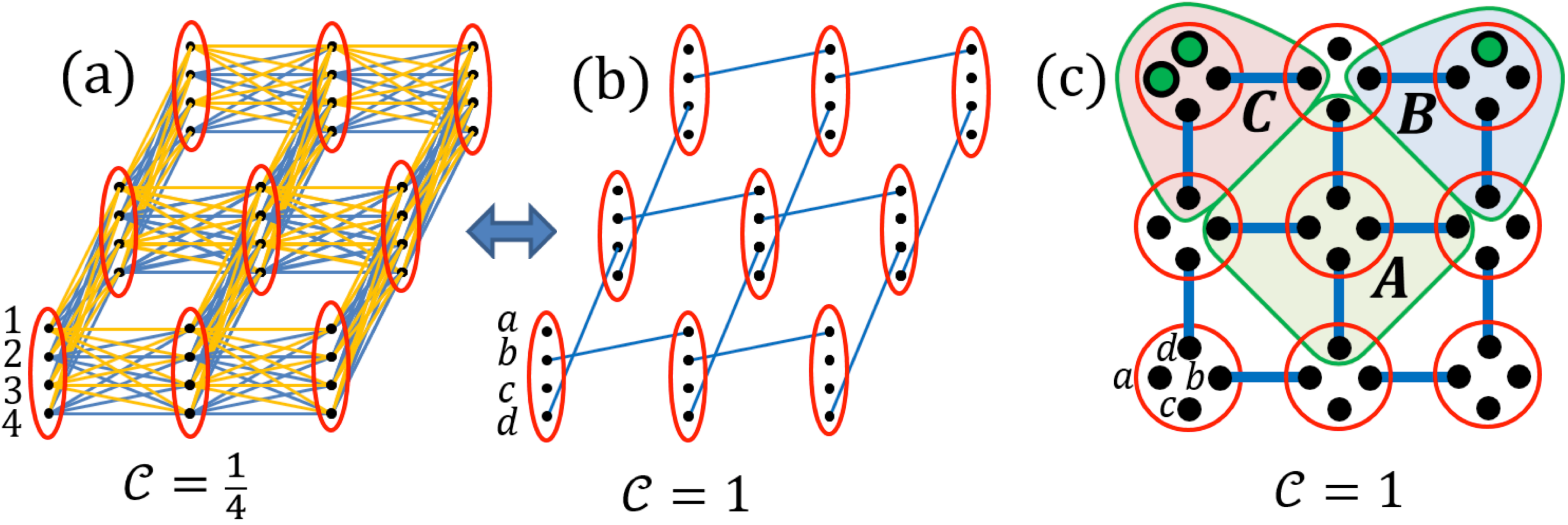}
\caption{\textbf{(a),~(b)}~Two equivalent ways to represent the quad-rail lattice %
using the simplified graphical calculus~\cite{Menicucci2011a}. The left graph represents the state using the physical-mode decomposition of each macronode, while the right graph represents the exact same state using the distributed-mode decomposition, with Eq.~\eqref{eq:physdist} connecting the two. 
Red ellipses indicate the macronodes (4-mode subsystems) that are left invariant by the change of mode decomposition.  \textbf{(c)}~Birds-eye view of the quad-rail lattice with respect to distributed modes with mode label conventions shown in the bottom left macronode. We include three input states and highlight three examples of input configurations within a macronode. In A, we have a ``blank" macronode that contains no input states. In B and C, respectively, one and two of the two-mode cluster states have been replaced with an input state. }
\label{fig:physdistrough}
\end{figure}

\section{Using macronodes for MBQC}\label{sec:macrointro}
The basic idea for our new protocol is that quantum computation can proceed via measurements on the QRL directly (rather than first reducing to the square-lattice cluster state). We break this section into five parts: encoding (Sec.~\ref{sec:encoding}), measurements (Sec.~\ref{sec:macromeas}), single-mode gates (Sec.~\ref{sec:smgates}), two-mode gates (Sec.~\ref{sec:2mgates}), and measurement readout (Sec.~\ref{sec:measreadout}).

\subsection{Encoding}
\label{sec:encoding}
%
%

In MBQC, once the resource state is prepared, the only allowable operations are local measurements. In our protocol, local measurements implement logic gates on \emph{macrolocally} encoded input states. This means that input states are localized with respect to a particular macronode, but they are distributed \emph{nonlocally} between the four physical modes that make it up. (The reason for this will become evident once we present our protocol.)

Each macronode admits two natural tensor-product decompositions. The first is the usual one defined in terms of the \emph{physical modes}~(P). The second---which is more useful for our purposes---is to define four \emph{distributed modes}~(D) as balanced linear combinations of the physical modes. Specifically, in the Heisenberg picture,
\begin{align}
\opvec{a}_{\text{D}}:=\mathbf{A}^{-1} \opvec{a}_{\text{P}}, \label{eq:physdist}
\end{align}
where $\opvec{a}_{\text{P}}:=(\op{a}_{1},\op{a}_{2},\op{a}_{3},\op{a}_{4})^\tp$ and $\opvec{a}_{\text{D}}:=(\op{a}_{a},\op{a}_{b},\op{a}_{c},\op{a}_{d})^\tp$. 
Note that numerical (alphabetical) subscripts are used for the physical (distributed) modes. 

The mapping in \eq{physdist} is exactly the inverse of a foursplitter gate [\eq{FS}]. Figure~\ref{fig:physdistrough} displays the QRL with respect to the physical modes~(a) and with respect to the distributed modes~(b). Notice that the former has fence-like connections between adjacent macronodes, while the latter consists merely of disjoint pairs. Also notice that the graphs in \fig{QRL}(a) and \fig{physdistrough}(b) are visually identical. Nevertheless, they represent different physical states because they are defined with respect to different mode decompositions (physical and distributed, respectively). 

For the rest of this Article, we will use distributed modes exclusively because this allows for the simplest description of information propagation through the QRL.
We allow input states to occupy any of the four possible distributed modes $(a, b, c, d)$ within a macronode. Unless otherwise specified, we assume that a maximum of two of the distributed modes within a given macronode are occupied by an input state. This guarantees that there is at least one two-mode cluster state per input that connects to an adjacent macronode. This condition is required in order to implement unitary gates (otherwise the output has no place to go). Three examples of input-state configurations are given in \fig{physdistrough}(c). 

%

\begin{figure}[t!]
\includegraphics[width=0.9\columnwidth]{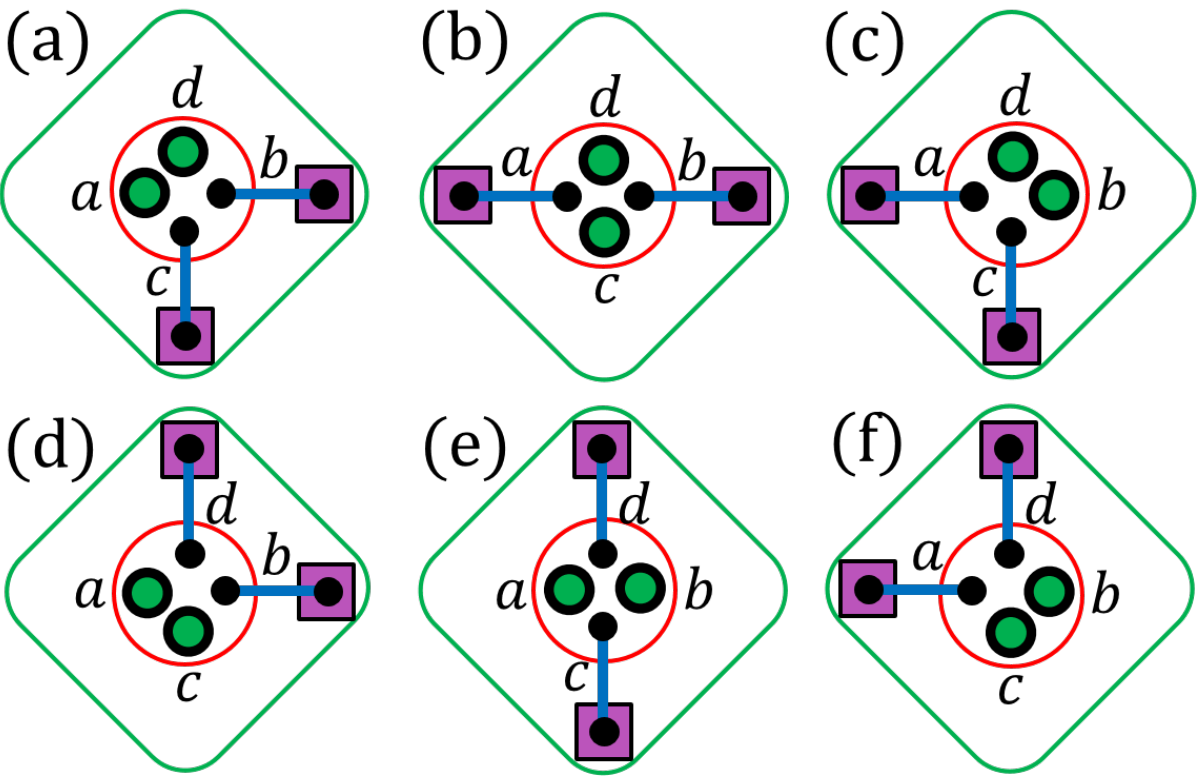}
\caption{Birds-eye view of six macronode configurations with two input states and two adjacent output modes (inside the purple boxes). 
We assign distributed-mode labels $\{a, b, c, d\}$ consistent with \fig{physdistrough}. Each input can be assigned to one of the two green modes, and each can also be mapped to either output, resulting in a total of 24 distinct processes. We omit macronode labels and labels on the output modes.} \label{fig:2inputconfigs}
\end{figure}

\subsection{Macronode measurements}
\label{sec:macromeas}
Our protocol implements Gaussian unitary gates on encoded input states by locally measuring the physical modes that make up each macronode in a rotated quadrature basis ${\hat{p}(\theta)\coloneqq \hat{p} \cos{\theta} - \hat{q} \sin{\theta}}$.
We vectorize the measurement bases for a given macronode measurement using 
\begin{align}
		\op{\mathbf{p}}_{\text{P}}(\boldsymbol\theta) & \coloneqq \begin{pmatrix} \hat{p}_{1} (\theta_{1}),   \hat{p}_{2} (\theta_{2}),   \hat{p}_{3} (\theta_{3}),   \hat{p}_{4} (\theta_{4}) \end{pmatrix}^\tp,  \label{eq:generalmeasure}
\end{align}
where $\boldsymbol\theta\coloneqq(\theta_{1}, \theta_{2}, \theta_{3}, \theta_{4})$. Note that local measurements with respect to the physical modes will generally correspond to nonlocal (four-body) measurements with respect to the distributed modes (and the inputs).

To characterize the effective logic gate implemented by macronode measurement, we consider the  two-input case (as in C in \fig{physdistrough}(c)). This case is the most general as the no- and single-input cases are special cases with both or one of the inputs replaced by half of a two-mode CV cluster state.

\begin{figure}[b!]
\includegraphics[width=1.0\columnwidth]{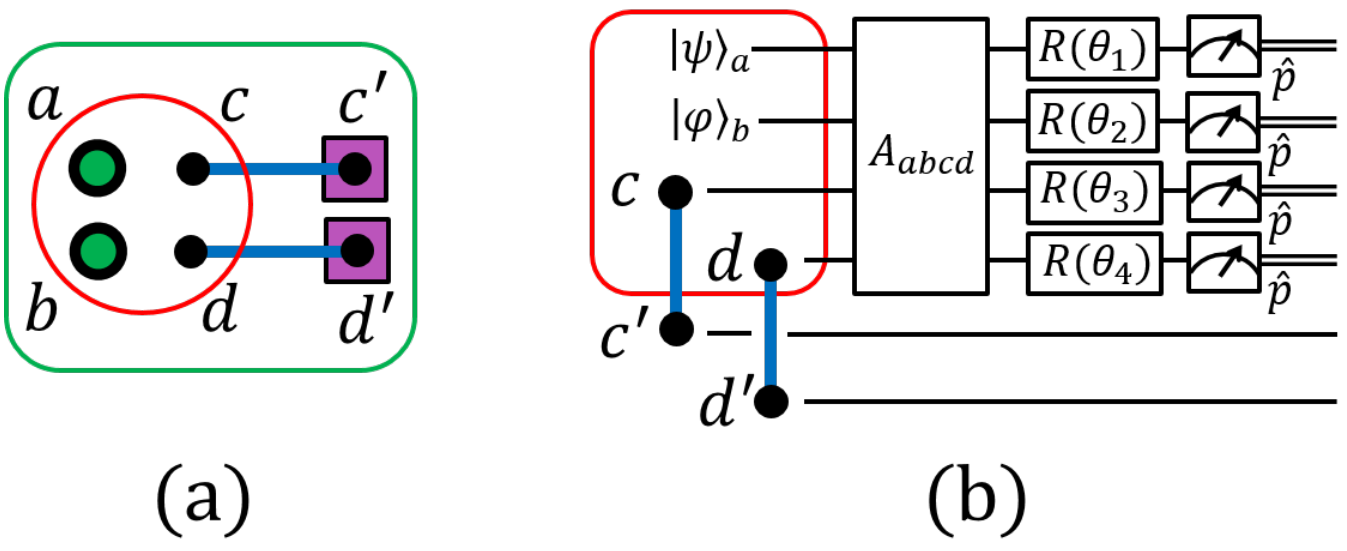}
\caption{\textbf{(a)}~A specific case of a single macronode with two input states and two adjacent output modes equivalent to Fig.~\ref{fig:2inputconfigs} (e). The output mode for $a$ is $c^{\prime}$ and the output mode for $b$ is $d^{\prime}$.  \textbf{(b)}~Quantum circuit for macronode measurement of a general two-input macronode for arbitrary inputs $\ket{\psi}$ and $\ket{\phi}$ encoded within distributed modes $a$ and $b$, respectively. Locally measuring the physical modes is exactly equivalent to first applying a foursplitter gate on the distributed modes and then doing the desired measurements. %
Up to measurement-dependent displacements and finite-squeezing effects, the output state is given in \eq{genevol}.} \label{fig:macromeasure2}
\end{figure}

There are ${\binom 4 2 = 6}$ different two-input macronode configurations (as shown in \fig{2inputconfigs}) and thus 12 total input configurations with distinct input states. In addition, each input must be paired with a two-mode cluster state that contains the corresponding output mode. There are two possibilities, resulting in 24 distinct input-to-output mode configurations. 
It suffices to characterize the single case shown in \fig{macromeasure2} because all other configurations are related to this by applying a permutation on the distributed modes prior to measurement, and this can be taken into account by a simple change to the homodyne angles.

\begin{figure*}[t!]
\includegraphics[width=1.8\columnwidth]{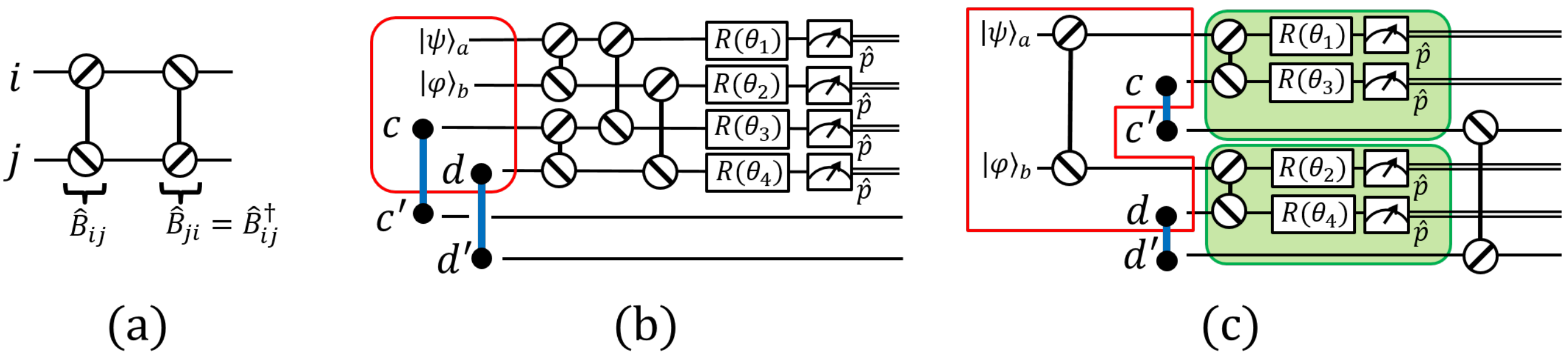}
\caption{\textbf{(a)}~Two-mode circuit representation of 50:50 beamsplitter $\hat{B}_{ij}$ and $\hat{B}_{ji}$. The overall effect of this circuit on modes $i$ and $j$ is to implement the identity gate. \textbf{(b)}~Macronode measurement as in \fig{macromeasure2} but with the four-splitter gate $\hat{A}_{abcd}$ replaced with four 50:50 beamsplitters, as in \eq{bsdecomp}. \textbf{(c)}~Restructured macronode measurement circuit equivalent to (b)  (see text for details). The vertical ordering of the modes has been changed. Modes enclosed within the red box belong to the measured macronode. Note that the two subcircuits within the green regions are identical up to the choice of measurement angles.  Each of these subcircuits can be individually interpreted as a CV teleportation protocol with generalized homodyne measurements~\cite{Braunstein1998}. Equivalently, they are each a single macronode measurement on the \emph{CV dual-rail quantum wire}, discussed in Ref.~\cite{Alexander2014}.}\label{fig:macromeasure3}
\end{figure*}

%
To see this, define a generic permutation gate via its four-mode symplectic matrix representation:
\begin{equation}
\label{eq:symplecticperm}
\boldsymbol{\sigma}= \begin{pmatrix} \tilde{\boldsymbol{\sigma}} & \mathbf{0} \\ \mathbf{0} & \tilde{\boldsymbol{\sigma}} \end{pmatrix}
\end{equation}
where $\tilde{\boldsymbol{\sigma}}$ is some $4\times 4$ permutation matrix (a single 1 entry in each row and column and all other entries 0).
It is sufficient to check the commutation properties of the four-splitter gate with each element of any generating set of all four-mode permutation gates.
Let $\boldsymbol\sigma_{jk}$ denote the permutation gate that swaps modes~$j$ and~$k$. Then we have that
\begin{align}
\mathbf{A}^{-1}\boldsymbol\sigma_{1,2}\mathbf{A}&=\boldsymbol\sigma_{2,4},\nonumber\\
\mathbf{A}^{-1}\boldsymbol\sigma_{1,3}\mathbf{A}&=\boldsymbol\sigma_{3,4},\nonumber\\
\mathbf{A}^{-1}\boldsymbol\sigma_{1,4}\mathbf{A}&=\boldsymbol\sigma_{2,3}\mathbf{R}_{2}(\pi)\mathbf{R}_{3}(\pi),
\end{align}
where $\mathbf{R}(\pi)$ is defined in Eq.~\eqref{eq:rotationgate}. Thus, by commuting through the four-splitter gate, each 4-mode permutation gate $\boldsymbol\sigma$ is mapped to a combination of a new permutation gates and some single-mode $\pi$ phase delays.
These gates can be incorporated directly into the macronode measurements by permuting the choice of measurement angles (e.g., $\theta_{i} \leftrightarrow \theta_{j}$) and adding $\pi$ phase delays (e.g., $\theta_{i} \mapsto \theta_{i} + \pi$).%

For the case shown in  \fig{macromeasure2} and neglecting measurement-dependent displacements and finite-squeezing effects (which are discussed in the proof below)\blk, the most general Gaussian unitary that can be applied on the two encoded input modes $\ket{\psi}$ and $\ket{\varphi}$ by measuring  in $\hat{\mathbf{p}}_{\text{P}}(\boldsymbol{\theta})$ is  
\begin{align}
	\ket{\psi}_{a}\ket{\varphi}_{b}\mapsto \hat{G}_{c^{\prime} d^{\prime}}(\vec\theta)\ket{\psi}_{c^{\prime}}\ket{\varphi}_{d^{\prime}}, \label{eq:genevol}
\end{align}
where $\theta_{1}\neq\theta_{3}$,  $\theta_{2}\neq\theta_{4}$, and
\begin{align}
	\op{G}_{jk}(\vec\theta):= \op{B}_{jk}^\dag \op{V}_{j}(\theta_{1}, \theta_{3}) \op{V}_{k}(\theta_{2},\theta_{4})\op{B}_{jk}. \label{eq:gengate}
\end{align} 
Sandwiched between the pair of 50:50 beamsplitters is the single-mode unitary gate
\begin{align}
&\op{V}_{j}(x, y)\coloneqq \nonumber \\
&\qquad \op{R}_{j}\left(\frac{x+y}{2}\right)\op{S}_{j}\left(\tan{\left[\frac{x-y}{2}\right]}\right)\op{R}_{j}\left(\frac{x+y}{2}\right). \label{eq:singlemodegate}
\end{align} 
Notice that the output states automatically emerge at distributed modes $(c',d')$ of adjacent macronodes.

\begin{proof}[Proof of Equation~\eqref{eq:genevol}]
We start with \fig{macromeasure3}, which shows a macronode measurement circuit where the  foursplitter is decomposed into four beamsplitters [using \eq{bsdecomp}]. To go from \fig{macromeasure3}(b) to \fig{macromeasure3}(c) we used an interferometric symmetry of the pair of two-mode cluster states on modes $(c, c^{\prime})$ and $(d, d^{\prime})$ derived in Appendix C of Ref.~\cite{Alexander2015}: acting with $\hat{B}_{c d}$ on this state is equivalent to acting with $\hat{B}_{d^{\prime} c^{\prime}}$ instead~\cite{Gabay2016}.

\fig{macromeasure3}(c) shows that macronode measurement is equivalent to two copies of a gate teleportation circuit~\cite{Alexander2014, Alexander2015} conjugated by beamsplitters ($\hat{B}_{a b}$ and $\hat{B}_{d^{\prime}c^{\prime}}$). The gate teleportation circuits each implement  
\begin{align}
\hat{V}(r, m_{j}, m_{k}, \theta_{j}, \theta_{k})\coloneqq\op{N}(r)\op{D} (m_{j}, m_{k}, \theta_{j}, \theta_{k}) \op{V}(\theta_{j}, \theta_{k})
\end{align}
where $j$ and $k$ are 1 and 3 (2 and 4) for the top (bottom) subcircuit in \fig{macromeasure3}(c), $\hat{V}$ is defined in \eq{singlemodegate}, and
\begin{align}
\op{D} (m_{j}, m_{k}, \theta_{j}, \theta_{k}) = \hat{D} \left[ \frac{-i e^{i\theta_{k}} m_{j} - i e ^{i\theta_{j}} m_{k}}{\sin(\theta_{j} - \theta_{k})} \right]
\end{align}
is a phase-space displacement [$\op{D} (\alpha )= e^{\alpha \hat{a}^{\dagger} - \alpha^{*} \hat{a}}$] that depends on the homodyne angles and measurement outcomes $m_{j}$ and $m_{k}$ associated with measuring modes $j$ and $k$.
Finally, 
\begin{align}
\hat{N}(r)=e^{- \varepsilon \hat{q}^{2} /2} e^{- \varepsilon \hat{p}^{2} /2 t^{2}} \hat{S}(t^{-1})
\end{align}
is a non-unitary operator that applies the noise from finite squeezing to the state  (after which the state must be renormalized%
)~\cite{Alexander2014}.  

The macronode measurement maps
\begin{align}
\ket{\psi}_{a}\ket{\varphi}_{b}\mapsto \hat{G}_{c^{\prime} d^{\prime}}(r, \mathbf{m}, \boldsymbol{\theta})\ket{\psi}_{c^{\prime}}\ket{\varphi}_{d^{\prime}}, \label{eq:genevolnoise}
\end{align}
where $\mathbf{m}=(m_{1}, m_{2}, m_{3}, m_{4})$, $\boldsymbol{\theta}=(\theta_{1}, \theta_{2}, \theta_{3}, \theta_{4})$, and
\begin{align}
\hat{G}_{j k}(r, \mathbf{m}, \boldsymbol{\theta})&\coloneqq \nonumber \\
\hat{B}_{k j}&\hat{V}_{j}(r, m_{1}, m_{3}, \theta_{1}, \theta_{3})\hat{V}_{k}(r, m_{2}, m_{4}, \theta_{2}, \theta_{4}) \hat{B}_{j k}.
\end{align}
In the limit of large squeezing and when all measurement outcomes are zero, we have:
\begin{align}
\hat{V}(\theta_{j}, \theta_{k})= \lim_{r\rightarrow\infty} \hat{V}(r, 0, 0, \theta_{j}, \theta_{k}), 
\end{align}
and so 
\begin{align}
\hat{G}(\boldsymbol{\theta})= \lim_{r\rightarrow\infty} \hat{G}(r, \mathbf{0}, \boldsymbol{\theta}).
\end{align}
In the more general case, the displacements can either be actively corrected at each step or merely accounted for using feedforward~\cite{Gu2009}.
From this, \eq{genevol} can be seen as the large squeezing limit of \eq{genevolnoise}. In the rest of this Article, we ignore displacements and finite-squeezing effects for simplicity of presentation.
\end{proof}

Note that for $\theta_{1}=\theta_{3}$ or $\theta_{2}=\theta_{4}$, \eq{singlemodegate} diverges in the squeezing factor and thus cannot represent a physical unitary operation. Nevertheless, the case where all four angles are equal ($\theta_{1}=\theta_{2}=\theta_{3}=\theta_{4}$) will later be shown to correspond to measurement readout; see Sec.~\ref{sec:measreadout}. Next we consider some examples of single- and two-mode Gaussian gates that are special cases of \eq{gengate}.

\subsection{Single-mode Gaussian unitary gates}
\label{sec:smgates}

%

The first examples we consider are single-mode Gaussian unitary gates. Consider restricting the homodyne angles so that 
\begin{align}
	 \theta_{2} &=\theta_{1} \qquad \text{and} \qquad \theta_{4}=\theta_{3}. 
	\label{eq:smthetarequirements}
\end{align}
In this case, the single-mode gates sandwiched between the beamsplitters above in \eq{gengate} are identical. Using \eq{BScom}, the beamsplitters cancel resulting in 
\begin{equation}
	\op{G}_{jk}(\boldsymbol\theta)\Bigr\rvert_{\substack{\theta_{2}=\theta_{1} \\ \theta_{4}=\theta_{3}}}=  \op{V}_{j}(\theta_{1}, \theta_{3}) \op{V}_{k}(\theta_{1},\theta_{3}), \label{eq:singlemoderestgate} 
\end{equation}
which implements a pair of single-mode gates on the input states. As the same gate gets implemented on both inputs, a single macronode measurement does not allow for the two input states to evolve independently. 

Independent single-mode gates can still be applied in the single-input case by ignoring the effect on the unused distributed mode. A single-mode $\hat{V}$ gate is sufficient to generate arbitrary single-mode Gausian unitary gates up to displacements (and only two applications are required for all of them)~\cite{Alexander2014}.

Applying further restrictions so that $\theta_{3}=\pm\theta_{1}$ implements a pair of phase delays and squeezers, respectively:
\begin{align}
	&\op{G}_{jk}(\boldsymbol\theta)\Bigr\rvert_{\theta_{4}=\theta_{3}= \theta_{2}=\theta_{1}}=\op{R}_{j}(2\theta_{1}) \op{R}_{k}(2\theta_{1}), 
\end{align}
and
\begin{equation}
	\op{G}_{jk}(\boldsymbol\theta)\Bigr\rvert_{\theta_{4}=\theta_{3}= -\theta_{2}=-\theta_{1}}=\op{S}_{j}(\tan \theta_1) \op{S}_{k}(\tan \theta_1).
\end{equation}
%

\subsection{Two-mode Gaussian unitary gates} \label{sec:2mgates} 
Here we provide different restrictions on the homodyne measurement angles $\boldsymbol\theta$ that yield interesting examples of two-mode gates from \eq{gengate}. 
Setting 
\begin{align}
	\theta_{3}=-\theta_{1} \qquad \text{and} \qquad
	\theta_{4}=-\theta_{2}
\end{align}
implements the two-mode-squeezing operation
\begin{align}
	\op{G}_{jk}(\boldsymbol\theta)\Bigr\rvert_{\substack{\theta_{3}=-\theta_{1} \\ \theta_{4}=-\theta_{2}}}= \op{B}_{jk}^\dag \op{S}_{j}(\tan \theta_{1}) \op{S}_{k}(\tan \theta_{2})\op{B}_{jk}.
\end{align}
We can also implement a linear-optics gate by setting 
\begin{align}
\theta_{3}=\theta_{1}-\frac{\pi}{2} \qquad %
\text {and} \qquad \theta_{4}=\theta_{2}-\frac{\pi}{2}.	
\end{align}
This implements
\begin{align}
&\hat{G}_{jk}(\boldsymbol\theta)\Bigr\rvert_{\substack{\theta_{3}=\theta_{1}-\frac{\pi}{2} \nonumber \\
		\theta_{4}=\theta_{2}-\frac{\pi}{2}}} \\
		&\quad = \op{B}_{jk}^\dag \op{R}_{j}\left(2\theta_{1}+\frac{\pi}{2}\right) \op{R}_{k}\left(2\theta_{2}+\frac{\pi}{2}\right)\op{B}_{jk}\nonumber \\ 
&\quad =\op{R}_{j}(\theta_{+}) \op{R}_{k}(\theta_{+}) \left[\op{R}_j\left(\frac{\pi}{2}\right) \op{B}_{jk}(\theta_{-})\op{R}_{k}\left(\frac{\pi}{2}\right) \right],
\end{align} 
where ${\theta_{\pm}=\theta_{1}\pm\theta_{2}}$. Thus, up to some additional phase delays, the above gate implements a variable beamsplitter.
%
\begin{figure}
\includegraphics[width=1\columnwidth]{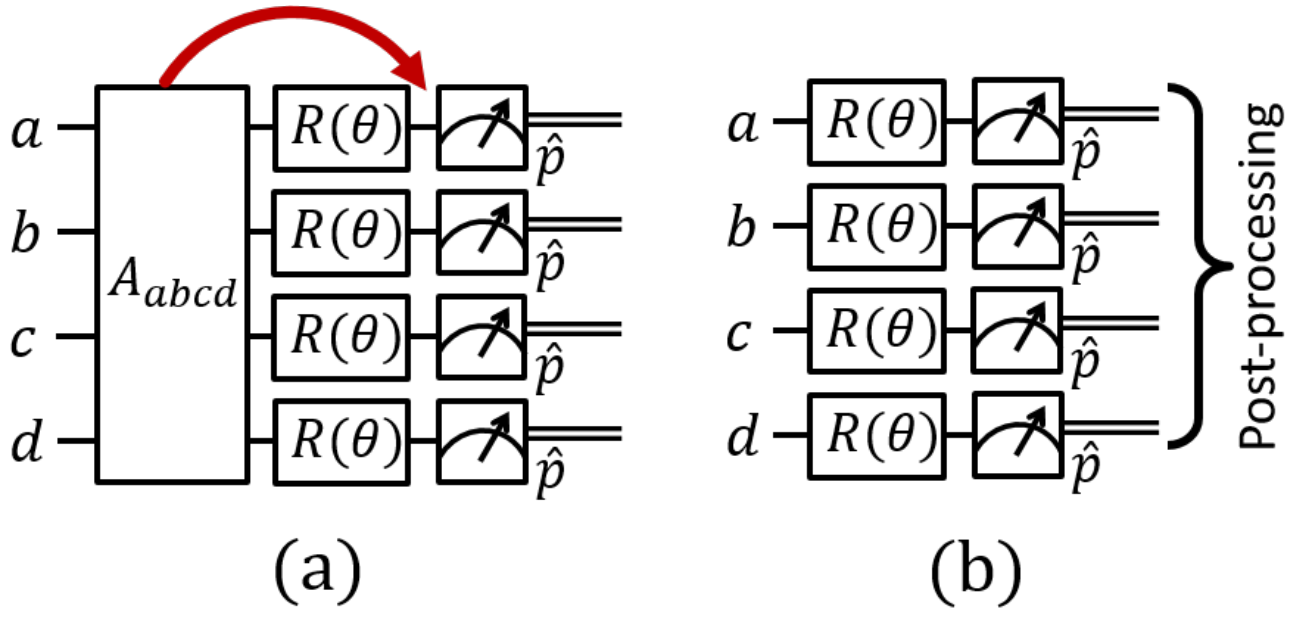}
\caption{\textbf{(a)}~Here we show a macronode measurement circuit with respect to the distributed modes. In the special case of restricting measurement angles such as in \eq{finalmeasure}, we can commute the phase delays past the foursplitter gate using \eq{com}, indicated by the red arrow. \textbf{(b)}~Acting with a foursplitter gate immediately before a collection of 
$\hat{p}$ measurements is equivalent to only measuring in $\hat{p}$ and then classically taking linear combinations of the outcomes (post-processing).}\label{fig:macromeasure4}
\end{figure}

\subsection{Measurement readout}\label{sec:measreadout}

In addition to implementing unitary gates, we must also be able to perform projective measurements on the encoded states.
This can be implemented directly on the QRL, and we allow up to (all) four of the distributed modes to be filled with inputs. Each input that shares a macronode during measurement readout will be measured in the same homodyne basis. (This means that modes to be measured in different bases must be located within different macronodes.) 

To measure each distributed mode within a single macronode in the homodyne basis  $\hat{p} (\theta)$, one simply has to apply the following restriction on the measurement angles:
\begin{align}
	\theta=\theta_{1}=\theta_{2}=\theta_{3}=\theta_{4}, \label{eq:finalmeasure}
\end{align}
as we now show.
By decomposing the foursplitter gate $\hat{A}$ using \eq{bsdecomp} and applying the beamsplitter commutation relations in \eq{BScom}, it is straightforward to verify that 
\begin{equation}
\left[\hat{A}_{ijkl}, \hat{R}_{i}(\theta)\hat{R}_{j}(\theta)\hat{R}_{k}(\theta)\hat{R}_{l}(\theta)\right]=0. \label{eq:com}
\end{equation}
Thus, with these restricted measurements, the foursplitter in the standard macronode measurement circuit as shown in \fig{macromeasure4}(a) can be commuted through the phase delays as shown. %

Measuring $\hat{p}$ on all physical modes after the gate $\hat{A}$ is equivalent to just measuring the modes in $\hat{p}$ and taking linear combinations (given by $\tilde{\mathbf{A}}$) of the measurement outcomes:
\begin{align}
\hat{A}_{1, 2, 3, 4}^{\dagger}\begin{pmatrix} \hat{p}_{1} \\ \hat{p}_{2} \\ \hat{p}_{3} \\ \hat{p}_{4} \end{pmatrix}\hat{A}_{1, 2, 3, 4} = \tilde{\mathbf{A}}\begin{pmatrix} \hat{p}_{1} \\ \hat{p}_{2} \\ \hat{p}_{3} \\ \hat{p}_{4} \end{pmatrix}.
\end{align}
The physical four-splitter that is applied can be undone by classical post-processing (applying~$\tilde{\mathbf{A}}^{-1}$) on the actual measurement outcomes. Thus, this macronode measurement can be implemented by measuring all of the distributed modes locally in the basis 
$\hat{p}(\theta)$%
, as shown in \fig{macromeasure4}(b).

\begin{figure}
\includegraphics[width=1\columnwidth]{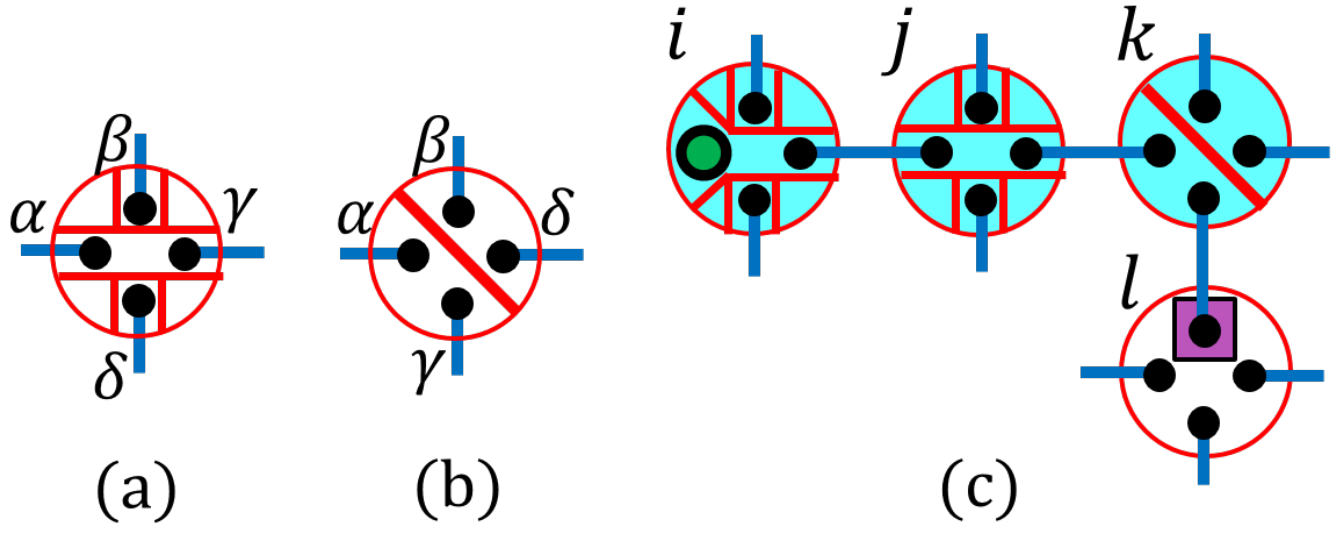}
\caption{\textbf{(a)}, \textbf{(b)}~Rather than labelling the modes within each macronode to indicate how inputs are mapped onto outputs, we introduce additional red lines that partition each macronode such that the pairs of labels $\{\alpha, \gamma\}$ and $\{\beta, \delta\}$ share a partition. (a) and (b) above show two examples of this. \textbf{(c)}~A connected sequence of macronodes $\{ i, j, k, l \}$ on the quad-rail lattice. Embedding of a quantum wire within the quad-rail lattice. We use light blue macronode coloring to indicate the use of the restricted measurements [as in \eq{smthetarequirements}].} \label{fig:wireembed}
\end{figure}

%

\begin{figure}
\includegraphics[width=1\columnwidth]{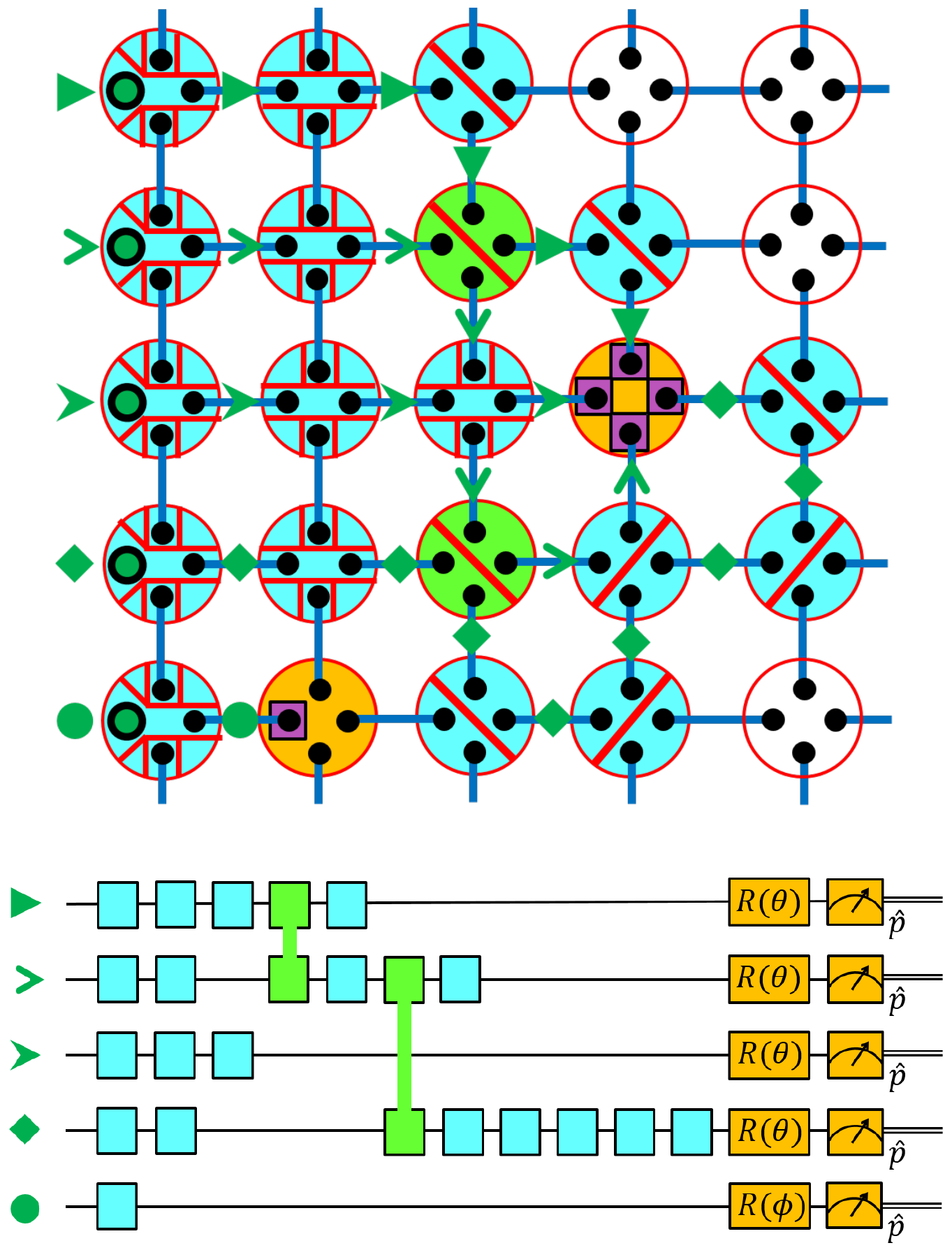}
\caption{\textbf{(Above)}~An example measurement scheme on the quad-rail lattice. There are five encoded input states on the left hand side, which we label by different arrow symbols. We label lattice edges to indicate how these inputs propagate along the lattice. Light blue macronode coloring indicates application of single-mode Gaussian unitaries only [of the form \eq{singlemoderestgate}]. Green macronode coloring indicates the application of a two-mode Gaussian unitary, such as those described in \sec{2mgates}. Orange macronode coloring is used to describe a measurement readout step on the lattice, as in \sec{measreadout}. \textbf{(Below)}~A quantum-circuit description of the overall Gaussian unitary implemented above. Light blue small boxes are single-mode Gaussian unitary gates and connected green boxes are two-mode Gaussian unitary gates. Measurement operations are colored orange. } \label{fig:flexcirc}
\end{figure}

\section{Constructing flexible quantum circuits}
\label{sec:flexcirc}

In the previous sections, we saw how input states can be encoded macrolocally [defined by \eq{physdist}] and how homodyne measurements on macronodes are sufficient to implement a variety of Gaussian unitary gates [of the form of \eq{gengate}], as well as  measurement readout. Now we briefly describe how connected regions of macronodes can be measured in order to implement quantum circuits.

We will start with how to construct quantum wires. In  Sec.~\ref{sec:smgates} we showed that for a specific configuration of input and output modes $(a\mapsto c^{\prime}, b\mapsto d^{\prime})$, restricting the measurement angles so that $\theta_{1}=\theta_{2}$ and $\theta_{3}=\theta_{4}$ ensures that the input states that share a macronode do not interact, i.e., only single-mode gates are applied. This result can be generalized for arbitrary input and output mode configurations by employing the 
permutation freedom discussed in \sec{macromeas}. By appropriately modifying the homodyne angles, we can apply the same single-mode gates and teleport inputs at sites $\alpha$ and $\beta$  to $\gamma^{\prime}$ and $ \delta^{\prime}$ respectively, for any valid assignment of  $\{ \alpha, \beta, \gamma, \delta \} \mapsto\{ a, b, c, d \}$. 
We represent this graphically as shown in \fig{wireembed}(a) and (b).

By restricting measurements like this on a connected sequence ${( i, j, k, \dotsc , l )}$ of macronodes on the QRL,  (up to displacements and finite-squeezing effects) we can implement a single-mode Gaussian unitary $\hat{V}_{l}\dotsm\hat{V}_{k}\hat{V}_{j}\hat{V}_{i}$ (omiting dependence on homodyne angles) on an input initially encoded within macronode $i$ and have it propagate through the sequence of macronodes ${( i, j, k, \dotsc , l )}$, outputing into macronode $l$. We illustrate this by way of example in \fig{wireembed}(c). These sequences thus act as embedded quantum wires, equivalent to the CV dual-rail wires described in Ref~\cite{Alexander2014}. 

Multiple wires can be embedded within the QRL provided that no two wires overlap on a lattice edge. Because we allow for up to two input states to share any macronode at a given time, these wires are free to intersect and cross one another. Note that when two wires meet at a macronode, the same single-mode Gaussian unitary gate gets applied to both inputs at that macronode. 

Alternatively, the macronodes that act as junctions between two wires can be used to implement a two-mode Gaussian unitary, as discussed in \sec{2mgates}. Therefore, wires and intersection sites can be used to implement single- and two-mode Gaussian unitary gates respectively, and these components are sufficient to generate arbitrary multi-mode Gaussian unitaries. Measurement readout (homodyne detection) can be implemented by connecting up to four wires to a given macronode and measuring it with restrictions as in \eq{finalmeasure}.

	By combining these results, we have a highly flexible means for implementing quantum circuits on the QRL. See \fig{flexcirc} for an example.  This is analogous to a field-programmable gate array (FPGA) since the QRL is a versatile resource that can be configured by the user at the ``software level''   into many different gate networks by the choice of measurement bases.  With access to vacuum input states and arbitrary displacements, these operations are sufficient to implement arbitrary Gaussian computations. 

%
\subsection*{Non-Gaussian resource}

Gaussian operations alone are known not to be universal for quantum computing~\cite{Bartlett2002}. Full universality can be achieved, however, by diverting a subset of the QRL nodes to photon counters instead of homodyne detectors~\cite{Gu2009,Menicucci2011a}. Depending on the particular practical implementation---which could even include encoded qubits and error correction~\cite{Menicucci2014}---it might be more favorable to periodically inject non-Gaussian resources~\cite{Wenger2004, Menicucci2014} instead of counting photons. %
We leave further discussion of such elements to future work.

\section{Comparison with previous work}
\label{sec:comparison}

How does our scheme compare with other previously established CV cluster-state protocols? Below, we compare it with three alternatives, focusing on the following four features: (1)~\emph{circuit flexibility}, which is the maneuverability of the quantum wires; (2)~\emph{compactness}, which is the minimum number of sites that must be measured in order to implement a desired class of gates; (3)~\emph{noise per gate due to finite squeezing}; and (4)~\emph{scalability}.

\emph{Canonical CV cluster state}.---The original CV measurement-based protocol introduced in Refs~\cite{Menicucci2006, Gu2009} uses a single-rail ${\mathcal{C}=1}$ square-lattice CV cluster state. Circuit flexibility is limited because the wires are generally constrained to run horizontally along the lattice, and two-mode gates can only be applied between nearest-neighbor wires. In general, single-mode Gaussian gates will require four steps along the lattice~\cite{Ukai2010,Alexander2014}, thus limiting compactness as well. 
The natural two-mode gate is limited to the $\hat{C}_{Z}$ gate. Noise due to finite squeezing is known to depend on the edge weight ($\mathcal{C}=1$)~\cite{Alexander2014}. As such, the amount of noise per single-mode Gaussian unitary gate is roughly similar between this protocol and the QRL protocol introduced here. %
This resource state is theoretically convenient to analyze, which is why it is often used for initial studies~\cite{Menicucci2006, Gu2009, Menicucci2014}, but it is less amenable to scalable design than macronode-based approaches (see Ref.~\cite{Alexander2015} and references therein).

\emph{Projected quad-rail lattice}.---The original CV measurement-based protocol can be modified to run on a ${\mathcal{C}=\tfrac 1 4}$ square-lattice cluster state~\cite{Alexander2014,Menicucci2011a}. This resource state has the advantage that it can be generated scalably (by the process shown in \fig{QRL}). This protocol has the same features as in the $\mathcal{C}=1$ case except with poorer noise properties~\cite{Alexander2014}. Specifically, the lower edge weight ${\mathcal{C}=\tfrac 1 4}$ means that using the QRL in this projected fashion will introduce significantly more noise (due to finite squeezing) than will applying the full QRL protocol introduced here.

\emph{Bilayer square lattice}.---We also consider the highly scalable bilayer-square-lattice (BSL) resource state recently introduced in Ref.~\cite{Alexander2015} (on which we are authors). Like the QRL, this state affords a similar macronode-based protocol, which we refer to here as the \emph{BSL protocol}. Like with the above two cases, circuit flexibility is limited because quantum wires are restricted to run horizontally, %
and the natural two-mode gates (which includes, but is not limited to, the $\hat{C}_{Z}$ gate) can only be applied between nearest-neighbor wires. In terms of compactness, the BSL protocol is similar to the QRL protocol since the individual wires themselves are actually CV dual-rail wires~\cite{Alexander2014}. For technical reasons, however, these wires require twice as many steps to implement each single-mode gate (four, as compared to the usual two). This results in poorer noise performance
than the QRL protocol.

Thus, our protocol shares the strengths of the others. It has relatively good noise performance (similar to the canonical CV cluster state), compactness (similar to the BSL) and scalability (similar to projected QRL and the BSL). In addition, it is the only protocol that allows highly flexible quantum circuit design: the extra degrees of freedom per site allow for the quantum wires to be more flexibly directed and even to criss-cross and intersect one another, thus simplifying two-mode interactions between initially distant wires.
In addition, the broad class of two-mode gates that can be implemented with a single macronode measurement include two-mode squeezing and a variable beamsplitter. Thus, the QRL protocol is especially well suited to quantum-optics applications.

\blk
\section{Conclusion}\label{sec:conc}
In this Article we generalized CV measurement-based protocols to a scalable cluster state known as the quad-rail lattice. This came with several advantages. In particular, we found that quantum wires can be threaded through the lattice sites, allowing for greater flexibility in implementing quantum circuits on the cluster. Unlike single-rail CV cluster-state wires~\cite{Gu2009}, these wires are embedded versions of the \emph{CV dual-rail wire} (discussed in Ref.~\cite{Alexander2014}), and thus, they are more compact and do not introduce excessive levels of noise due to finite squeezing~\cite{Alexander2014}. 
Our protocol is also well suited to implementing  a variety of two-mode gates at the intersection points of these wires---such as two-mode squeezing and beamsplitter gates. Thus, we have generalized the one-dimensional macronode protocols introduced in Ref.~\cite{Alexander2014} to the two-dimensional case.

Several novel features that our protocol exhibits---including nonlocal input states and the ability to re-route wires---are similar to those found in  generalizations of measurement-based quantum computing based on tensor networks~\cite{Gross2007a, Gross2007}. These similarities likely stem from their shared use of entangled pairs as basic building blocks. It is curious that these extra features are naturally exhibited in experimentally favorable schemes for implementing CV cluster state computations. It is worth considering the possibility that macronode-based qubit resource states might show similar advantages.

This work highlights the importance of focusing on macronode-based construction methods of CV resource states for quantum computing~\cite{Menicucci2011a, Yokoyama2013, Wang2013, Chen2014, Alexander2015}, which also have the advantage of being the most scalable methods available to date. Adapting the measurement-protocol to the quad-rail lattice---rather than converting it to the standard square-lattice resource---yields a richer, more dynamic mode of computation and opens further research avenues towards closing the gap between theoretical models and experimental implementations.

\acknowledgments
 We thank Peter Van Loock and Cameron Duncan for useful discussions. This work was supported by the Australian Research Council under Grant No. DE120102204 and by the U.S. Defense Advanced Research Projects Agency (DARPA) Quiness program..

\bibliographystyle{jacob_bibstyle}
\bibliography{ref,MenicucciPapersRefs}

\end{document}